\documentstyle[aps,preprint]{revtex}
\begin{document}
\draft
\preprint{ISSP December}
\title{Superconducting correlation
in the one-dimensional $t$-$J$ model
with
anisotropic spin interaction
and
broken parity}
\author{J. Shiraishi, Y. Morita and
M. Kohmoto}
\address{Institute for Solid State Physics,
 University of Tokyo
 7-22-1 Roppongi Minato-ku,\\ Tokyo 106, Japan}
\maketitle
\begin{abstract}
A variant of the one-dimensional
$t$-$J$ model with
anisotropic spin interaction
and
broken parity
is studied
by
the nested algebraic Bethe-ansatz method.
The gapless charge excitations
and the gapful spin excitations are obtained.
It is shown that the singlet-superconducting correlation
dominates in the low-density region
by applying
the finite-size scaling analysis in the conformal field theory.
\end{abstract}
\narrowtext
In these years,
strongly-correlated electron systems have drawn much attention.
It is partly caused by the discovery
of the copper-oxide high-$T_c$ superconductors.
The one-dimensional models of strong correlation play an
important role since some of them can be solved exactly.
The non-perturbative results thus obtained
contain some of the essential
properties of the strongly correlated systems.
Close scrutiny, however, is required when one tries to discuss
higher dimensions based on the one-dimensional results,
since some of them are specific to one dimension.

Besides the Hubbard model\cite{h1,h2,h3},
the $t$-$J$ model \cite{tj}
is
regarded as one of the most basic models
which contains the essence of strong correlation.
The Hamiltonian is
$
{\cal H}_{tJ}
=\sum _{<i,j>}\left[
 -t\; \sum _{{\sigma=\uparrow,\downarrow }}
 {\cal P} (c_{i {\sigma }}^{\dagger }c_{j {\sigma }}
     +c_{j {\sigma }}^{\dagger }c_{i {\sigma }}){\cal P}
 +
J\;
(S_{i}^{x}S_{j}^{x}+S_{i}^{y}S_{j}^{y}+S_{i}^{z}S_{j}^{z}
-{\frac {1}{4}}\ n_{i}n_{j})
\right],
$
where $n$'s are the number operators given by
$n_i=n_{i\uparrow}+n_{i\downarrow}=
c^{\dag}_{i\uparrow}c_{i\uparrow}+
c^{\dag}_{i\downarrow}c_{i\downarrow}$ and
the spin operators are
$S^k_i=\frac{1}{2}\sum_{\alpha,\beta}
 \;c_{i \alpha}^{\dagger }\sigma^k_{\alpha,\beta}c_{i \beta}$
with the usual Pauli matrices $\sigma$'s.
The Gutzwiller projector
${\cal P}={\prod }_{j=1}^{L}(1-n_{j {\uparrow }}n_{j {\downarrow
    }})$
restricts the Hilbert space by forbidding
double occupancies hence represents strong correlation.
In one dimension,
the exact solution was obtained
for the supersymmetric case
($2t=J$)\cite{suth,schlot,bares1,bares2}
using the Bethe-ansatz method.
The long-distance behavior
of the correlation functions was also investigated
by applying
the finite-size scaling analysis
in the conformal field theory
to the excitation spectra obtained by the Bethe-ansatz method\cite{ky}.
It was shown that
the superconducting correlation
does not exceed the others
such as the spin density wave (SDW) and the charge density wave (CDW)
correlations
at any filling.
However, the region
where the superconducting correlation is dominant
was found between the low-density supersymmetric region and
the ``phase-separated'' region ($J\gg t$)
by
numerical diagonalization of finite clusters\cite{tjpd}.
Since the numerical results are not totally reliable
due to the finite-size effect,
it is highly desirable to have exact results.

In this paper
we study
a one-parameter family
of the one-dimensional correlated electron systems
which includes
the ordinary supersymmetric $t$-$J$ model.
The Hamiltonian is
\begin{eqnarray}
{\cal H}_{tJ}
&=&\sum_{i}\Biggl[
 -\; \sum _{{\sigma }}
 {\cal P} (c_{i {\sigma }}^{\dagger }c_{i+1 {\sigma }}
     +c_{i+1 {\sigma }}^{\dagger }c_{i {\sigma }}){\cal P}
 +
2\;  (S_{i}^{x}S_{i+1}^{x}+S_{i}^{y}S_{i+1}^{y}+{\Delta}
S_{i}^{z}S_{i+1}^{z}
-\frac{\Delta}{4} n_{i}n_{i+1}
)
\Biggr.\nonumber\\
\Biggr.
&&  +
\eta\; (S_{i}^{z}n_{i+1}-n_{i}S_{i+1}^{z})
 + 2\Delta\; n_i \Biggr],
\label{ham1}
\end{eqnarray}
where
${\Delta }^{2}-{\eta }^{2}=1$ and we
parameterize them as
$
\Delta=\cosh{\gamma}$ and $\eta=\sinh{\gamma} \;\;(\gamma \in {\bf
  R_{\geq 0}})
$\cite{XYham}.
When $\gamma=0$, our Hamiltonian reduces to the
ordinary supersymmetric
$t$-$J$ model.
When the number of electrons
coincides with
the number of lattice sites,
our model becomes
$s={\frac {1}{2}}$ XXZ spin chain.
The Hamiltonian (\ref{ham1}) commutes with the
transfer matrix of the
solvable two-dimensional classical lattice model
associated with the supersymmetric quantum affine superalgebra
$U_q(\widehat{sl(2|1)})\ (q=e^{\gamma })$, which is a special case of the
model given by Perk and Schultz\cite{ps,devega}.
Thus one could say
that the Hamiltonian is ``$q$-supersymmetric''.
The ground state, excitations
and the correlation functions
are obtained by the nested algebraic Bethe-ansatz method,
the finite-size scaling analysis in the conformal field theory and
the numerical diagonalizaton for small clusters.
We find spin gap
and observe that the superconducting correlation dominates in the
low-density region in contrast to
the ordinary supersymmetric $t$-$J$ model.

The third term of the Hamiltonian ($\ref{ham1}$)
breaks the parity invariance.
The role of this term
was investigated
by
diagonalizing the system without it numerically.
It turned out that the spin gap survives in some cases.

{}~\vspace{3mm}

{\bf 1}. {\sl Bethe-Ansatz Equations (BAEs).}\ \
The diagonalization of the Hamiltonian (\ref{ham1})
with periodic boundary condition
reduces to solving the coupled algebraic equations (BAEs)
derived by the nested algebraic Bethe-ansatz technique.
They are
\begin{eqnarray}
{\biggl (}\frac {\sin (p_{j}+\frac {i}{2} \gamma )}
{\sin (p_{j}-\frac {i}{2} \gamma )}{\biggr )}^{L}
=(-1)^{N}\prod _{\beta =1}^{M}
\frac {\sin (p_{j}-\Lambda _{\beta }+\frac {i}{2} \gamma )}
{\sin (p_{j}-\Lambda_{\beta }-\frac {i}{2} \gamma )}\;\;\;\;\;
j=1, 2, \cdots ,N,
\label {bae1}
\\
\prod _{j =1}^{N}
\frac {\sin (\Lambda _{\alpha }-p_{j}+\frac {i}{2} \gamma )}
{\sin ( \Lambda _{\alpha }-p_{j}-\frac {i}{2} \gamma )}
=-\prod _{\beta =1}^{M}
\frac {\sin (\Lambda _{\alpha }-\Lambda _{\beta }
                        +i \gamma )}
{\sin ( \Lambda _{\alpha }-\Lambda_{\beta }-i \gamma )}\;\;\;\;\;
{\alpha }=1, 2, \cdots ,M,
\label {bae2}
\end {eqnarray}
where
$L$ is the number of lattice sites,
$N$ is the number of electrons,
$M$ is the number of down-electrons (magnons),
$p$'s are the quasi-momenta of electrons
and $\Lambda $'s are the magnon rapidities \cite {full}.

Hereafter we take the following ansatz
for the ground state and the elementary excitations:
{\it $\Lambda $'s are one-strings $\{\Lambda_{\alpha}\in {\bf R}|
\alpha=1,\cdots,M\}$
and
$p$'s consist of
the one-strings
$\{p_j=u_j\in {\bf R}| j=1,{\cdots },N-2M\}$
and
the two-strings
$\{
p_{{\alpha }}^{\pm}=\Lambda_{{\alpha }}{\pm }i\gamma/2|
{\alpha }=1,{\cdots },M\}
$.}
Note that the real parts of the two-strings coincide with the
magnon rapidities and, if $M=N/2$, there is no degrees of freedom
for one strings of the quasi-momenta.
This is essentially the same
ansatz established in Refs.{\cite {schlot,bares1,bares2}}.
In our case, however,
$\Lambda$'s and $u$'s
are in the interval  $[-{\frac {\pi }{2}}, {\frac {\pi }{2}}]$
due to the periodicity of the BAEs (\ref{bae1}), (\ref{bae2}).
By taking the logarithm of the BAEs,
we have
\begin{eqnarray}
&L\; \phi(u_j,\frac{\gamma}{2})=&2 \pi i I_j +
        \sum_{\beta=1}^M \phi(u_j-\Lambda_{\beta},\frac {\gamma }{2})
\qquad\qquad j=1,\cdots,N-2M,
\label{eq:BAE1}\\
&L \;\phi(\Lambda_{\alpha},\gamma)=&2 \pi i J_{\alpha} +
 \sum_{j=1}^{N-2M} \phi(\Lambda_{\alpha}-u_j,\frac{\gamma}{2})
+\sum_{\beta=1}^M \phi(\Lambda_{\alpha}-\Lambda_{\beta},\gamma)
\qquad \alpha=1,\cdots,M,
\label{eq:BAE2}
\end{eqnarray}
where
$\phi (z,{\alpha })
\equiv\log \frac {\sin (z+i {\alpha })}{\sin (z-i {\alpha })}$
(see \cite {branch}),
and
$\{\ I_j\ |\ j=1, 2, \cdots,N-2M\}$ is a set of integers
(or half-odd integers)
if $M$ is even (or odd) and the set
$\{\ J_{\alpha}\ |\ \alpha=1, 2, \cdots,M\}$
is a set of integers (or half-odd integers)
if $N+M+1$ is even (or odd).
We order the quantum numbers $I$'s and $J$'s
according to $I_j>I_{j+1}$ and $J_{\alpha}>J_{\alpha+1}$.

In the
thermodynamic limit
$L,N\rightarrow \infty$,
the distributions of $\Lambda$'s can be described
by the continuous density given by
$
L\; {\rho }(\Lambda_{\alpha})=\lim_{L,N\rightarrow \infty}
1/(\ \Lambda_{\alpha+1}-\Lambda_{\alpha}\ ).
$
The energy and momentum up to $O(1)$ are
\begin{equation}
E
=i\ {\sinh {\gamma }} \;
{\sum _{j=1}^{N}}
 {\phi }^{\prime}(p_{j},{\frac {\gamma }{2}})
=i\ {\sinh {\gamma }} \left(
L\int d\Lambda' \;\rho(\Lambda')\;
 {\phi }^{\prime}(\Lambda',\gamma)
+
{\sum _{j=1}^{N-2M}}
 {\phi }^{\prime}(u_{j},{\frac {\gamma }{2}})
\right),
\label {e}
\end{equation}
and
\begin{equation}
P
=i\
{\sum _{j=1}^{N}}
 {\phi }(p_{j},{\frac {\gamma }{2}})
=
i\
L\int d\Lambda' \;\rho(\Lambda')\;
 {\phi }(\Lambda',\gamma)
+
i\
{\sum _{j=1}^{N-2M}}
 {\phi }(u_{j},{\frac {\gamma }{2}}).
\label {p}
\end{equation}
The integral intervals in (\ref{e}) and (\ref{p})
will be discussed
in the following sections.

{}~\vspace{3mm}

{\bf 2}. {\sl Ground State.}\ \
We set
$N$ to be even for simplicity.
Since the spin interaction is anti-ferromagnetic,
the total $S^z$ for the ground state is expected to be zero.
This can be achieved by setting $M=N/2$, i.e.
for the sector without one-strings $p_j=u_j$.
We also require that the momentum $P$ to be zero.
We propose the following ansatz for $J$'s:
{\it
the distributions of
$J$'s for the ground state
is restricted as
$J_{\rm max}\geq |J_{\alpha}|\geq J_{\rm min}$, where
$J_{\rm max}=\frac{L-M-1}{2}$ and
$J_{\rm min}=\frac{L-2M+1}{2}$.}

In the thermodynamic limit,
we assume that $\Lambda$'s are distributed only in the regions
$[-{\pi }/2,-Q_{g}]$ and $[Q_{g},{\pi }/2]$ in accordance with
the distribution of the quantum numbers $J$'s.
BAEs (\ref{eq:BAE1}) and (\ref{eq:BAE2})
are reduced to
\begin{equation}
2 {\pi }i\;{\rho }_{g}(\Lambda)
=
-\phi ^{\prime }(\Lambda,\gamma )
+
{\bigg [}{\int}_{-{\pi }/2}^{-Q_{g}}+{\int}_{Q_{g}}^{{\pi }/2}{\bigg ]}
d\Lambda' \;
{{\rho }_{g}(\Lambda' )}\;\phi ^{\prime }(\Lambda-\Lambda',{\gamma }),
\label{intg}
\end{equation}
where $Q_{g}$ is determined by
$
\left[ {\int}_{-{\pi }/2}^{-Q_{g}}+{\int}_{Q_{g}}^{{\pi }/2}\right]
d\Lambda \;{{\rho }_{g}(\Lambda )}={\frac {N/2}{L}}\equiv \frac{n}{2}.
$
The ground-state energy $E_{g}$ is given by
\begin{equation}
E_{g}=i\ L\;{\sinh {\gamma }}\
{\bigg [}{\int}_{-{\pi }/2}^{-Q_{g}}+{\int}_{Q_{g}}^{{\pi }/2}{\bigg ]}
d\Lambda \;{{\rho }_{g}(\Lambda )}\;\phi ^{\prime }({\Lambda },{\gamma }).
\label{eg}
\end{equation}
These equations
can be solved numerically for arbitrary filling $n$.
The results are given in Fig. 1.
{}~\vspace{3mm}

{\bf 3}. {\sl Charge Excitations.}\ \
The charge excitations are those caused by
the replacements of the $\Lambda$'s
while keeping $M=N/2$
namely $S^{z}$ remains to be zero.
Thus {\it the elementary excitations for the charge sector
consists in making a jump (hole) at the point $J_{\alpha_{h} }$
and putting a quantum number $J_{\alpha_p}$at
a previously unoccupied region}{\cite{bares1,bares2}}.
In the thermodynamic limit,
BAEs (\ref{eq:BAE1}) and (\ref{eq:BAE2})
are reduced to
\begin{eqnarray}
2 {\pi }i\;{\rho }_c(\Lambda)
&=&
-\phi ^{\prime }(\Lambda,\gamma )
-{\frac {2{\pi }i}{L}}\;{\delta }(\Lambda-\Lambda_{h})
+{\frac {1}{L}}\phi ^{\prime }(\Lambda-\Lambda_{p},
 {\gamma })\nonumber\\
&&+
{\bigg [}{\int}_{-{\pi }/2}^{-Q_c}+{\int}_{Q_c}^{{\pi }/2} {\bigg ]}
d\Lambda' \;{{\rho }_c(\Lambda' )}\;
 \phi ^{\prime }(\Lambda-\Lambda',{\gamma }),
\label{intc}
\end{eqnarray}
retaining terms up to $O(L^{-1})$,
where $\Lambda_{p}$ and $\Lambda_{h}$ denote
the position of the hole and particle
in the sea of two strings
associated with the quantum number $J_{\alpha_h}$ and $J_{\alpha_p}$
respectively.
$Q_c$ is determined by
$
\left[{\int}_{-{\pi }/2}^{-Q_c}+{\int}_{Q_c}^{{\pi }/2}\right]
d\Lambda \;{{\rho }_c(\Lambda )}={\frac {(N-2)/2}{L}}.
$
For convenience,
we decompose ${\rho }_c({\Lambda })$
into the regular part and the singular part as
$
{\rho }_c(\Lambda)
={\rho }_{c0}(\Lambda)
-{\frac {1}{L}}{\rho _{c1}}(\Lambda)
-{\frac {1}{L}}{\delta }(\Lambda-\Lambda_{h}),
$
where ${\rho }_{c0}(\Lambda)$ satisfies
\begin{equation}
2 {\pi }i\;{\rho }_{c0}(\Lambda)
=
-\phi ^{\prime }(\Lambda,\gamma )+
{\bigg [}{\int}_{-{\pi }/2}^{-Q_c}+{\int}_{Q_c}^{{\pi }/2}{\bigg ]}
d\Lambda' \;{{\rho }_{c0}(\Lambda' )}\;
\phi ^{\prime }(\Lambda-\Lambda',{\gamma }).
\label{intg2}
\end{equation}
For $\rho _{c1}(\Lambda)$, we have
\begin{equation}
2 \pi i\;\rho _{c1}(\Lambda)
=\phi ^{\prime }(\Lambda-\Lambda_{h},{\gamma })
-\phi ^{\prime }(\Lambda-\Lambda_{p},{\gamma })
+
{\bigg [}{\int}_{-{\pi }/2}^{-Q_c}+{\int}_{Q_c}^{{\pi }/2}{\bigg ]}
d\Lambda' \;{{\rho }_{c1}(\Lambda' )}\;
\phi ^{\prime }(\Lambda-\Lambda',{\gamma })
\label{cint2}
\end{equation}
The excitation energy ${\Delta }E$ from the ground state
and the momentum $P$
are given by
\begin{equation}
{\Delta }E
=
i\ {\sinh }{\gamma }
\left( \phi ^{\prime }(\Lambda_{p},{\gamma })
 -\phi ^{\prime }(\Lambda_{h},{\gamma })
-\left[{\int}_{-{\pi }/2}^{-Q_c}+{\int}_{Q_c}^{{\pi }/2}\right]
d\Lambda \;{{\rho }_{c1}(\Lambda )}\;
\phi ^{\prime }({\Lambda },{\gamma })\right),
\label{ce}
\end{equation}
\begin{equation}
P
=
i\
\left( \phi (\Lambda_{p},{\gamma })
 -\phi (\Lambda_{h},{\gamma })
-\left[{\int}_{-{\pi }/2}^{-Q_c}+{\int}_{Q_c}^{{\pi }/2}\right]
d\Lambda\; {{\rho }_{c1}(\Lambda )}\;\phi ({\Lambda },{\gamma })\right).
\label{cp}
\end{equation}
Solving $(\ref {cint2})$ numerically,
the dispersion relation for the elementary charge excitations
was obtained.
The result for $\gamma=2$ and $n=0.45$ is shown in Fig. 2.
The results for other parameters
do not change in an essential manner, namely
{\it the charge excitation is always gapless}.

{}~\vspace{3mm}

{\bf 4}. {\sl Spin Excitations.}\ \
The spin excitations can be considered as excitations
coming from destroying the two-strings $p^{\pm}$'s and creating
one-strings $u$'s.
To study the elementary ones, let us consider
the case of $M=N/2-1$ magnons.
We assume that,
{\it in the sea of the quantum numbers $J$'s
there are no jumps}{\cite {bares1,bares2}}.
Then, in the thermodynamic limit, BAE (\ref{eq:BAE2})
becomes
\begin{eqnarray}
2 \pi i \; \rho_s (\Lambda)
&=&
-
\phi ^{\prime }(\Lambda,{\gamma })+
{\frac {1}{L}}\phi ^{\prime }(\Lambda-u_{1},{\frac {\gamma }{2}})+
{\frac {1}{L}}\phi ^{\prime }(\Lambda-u_{2},{\frac {\gamma }{2}})
\nonumber\\
&&+
{\bigg [}{\int}_{-{\pi }/2}^{-Q_s}+{\int}_{Q_s}^{{\pi }/2}{\bigg ]}
d\Lambda'\;{{\rho_s }
({\Lambda' })}\;\phi ^{\prime }(\Lambda-\Lambda',{\gamma }),
\label{rsint}
\end{eqnarray}
where
$u_{1}$ and $u_{2}$ are one-string quasi-momenta, and
$Q_s$ is determined by
$
\left[{\int}_{-{\pi }/2}^{-Q_s}+{\int}_{Q_s}^{{\pi }/2}\right]
d{\Lambda }{{\rho_s }({\Lambda })}={\frac {(N-2)/2}{L}}.
$
It is convenient to decompose ${\rho_s }(\Lambda )$
into contributions of order $O(1)$ and $O(L^{-1})$ as
$
{\rho_s }({\Lambda })={\rho }_{s0}({\Lambda })
-{\frac {1}{L}}{\rho }_{s1}({\Lambda }),
$
where ${\rho }_{s0}(\Lambda )$ satisfies the same equation as
({\ref  {intg2}}) obtained by replacing all the suffices $c$ to $s$.
Then the integral equation for ${\rho }_{s1}({\Lambda })$
is obtained as
\begin{equation}
2 \pi i\; {\rho }_{s1}(\Lambda)=
-\phi ^{\prime }(\Lambda-u_{1},
{\frac {\gamma }{2}})
-\phi ^{\prime }(\Lambda-u_{2},
{\frac {\gamma }{2}})
+
{\bigg [}{\int}_{-{\pi }/2}^{-Q_s}+{\int}_{Q_s}^{{\pi }/2}{\bigg ]}
d\Lambda'{{\rho }_{s1}(\Lambda')}\phi ^{\prime }(\Lambda-\Lambda',{\gamma }),
\label{rsint2}
\end{equation}
The excitation energy ${\Delta }E$ from the ground state
and the momentum $P$ are
\begin{equation}
{\Delta }E
=
i \ {\sinh }{\gamma }\;
\left( \phi ^{\prime }(u_{1},{\frac {\gamma }{2}})
 +\phi ^{\prime }(u_{2},{\frac {\gamma }{2}})
-\left[{\int}_{-{\pi }/2}^{-Q_s}+{\int}_{Q_s}^{{\pi }/2}\right]
d\Lambda \;{{\rho }_{s1}(\Lambda)}\;
\phi ^{\prime }({\Lambda },{\gamma }) \right)
\label{se}
\end{equation}
\begin{equation}
P
=
i
\left( \phi (u_{1},{\frac {\gamma }{2}})
 +\phi (u_{2},{\frac {\gamma }{2}})
-\left[{\int}_{-{\pi }/2}^{-Q_s}+{\int}_{Q_s}^{{\pi }/2}\right]
d\Lambda \;{{\rho }_{s1}(\Lambda )}\;\phi ({\Lambda},{\gamma }) \right).
\label{sp}
\end{equation}
Solving $(\ref {rsint2})$ numerically,
the dispersion relation for the elementary spin excitations
was obtained.
The result for $\gamma =2$ and $n=0.45$
is shown in Fig. 3.
The results for other parameters
do not change in an essential manner, namely
{\it the spin excitation is always gapful}.
The spin gap as a function of $\gamma $ is also shown in Fig. 4.
and one can see that the gap increases as holes are doped.

To study the effect of the parity violating term,
we calculated
the spin-spin correlations $\langle S^z_i  S^z_j\rangle $
for the parity-unbroken Hamiltonian
${\cal H}_{tJ}-\sum_i \eta(S^z_i n_{i+1}-n_i S^z_{i+1})$\cite {pru}.
The results are shown in Fig. 5
and they indicate that
the correlation decays exponentially.
Hence there is still excitation gap.

{}~\vspace{3mm}

{\bf 5}. {\sl Correlation Functions.}\ \
Consider a field-theoretic description of
the low-lying excitations.
Since the dispersion
for the low energy charge sector is
approximately linear for $0<n<1$,
and the gapful spin sector is
irrelevant for the low-energy behavior,
we can expect the system can be described
by the conformal field theory \cite{bpz}.

Let us consider the excitations described by
the density $\rho(\Lambda)$ satisfying
\begin{equation}
2 {\pi }i\;{\rho }(\Lambda)
=
-\phi ^{\prime }(\Lambda,\gamma )
+
{\bigg [}{\int}_{-{\pi }/2}^{Q_{-}}+{\int}_{Q_{+}}^{{\pi }/2}{\bigg ]}
d\Lambda' \;{{\rho }(\Lambda' )}\;
\phi ^{\prime }(\Lambda-\Lambda',{\gamma }),
\label{eq:KY1}
\end{equation}
and
apply
the general method
of Kawakami-Yang \cite{ky}
for the
finite-size scaling method \cite{af,car}.
Using the Fourier-transform technique,
we rewrite (\ref{eq:KY1}) as
\begin{equation}
{\rho }(\Lambda)
=
2R_q(2\Lambda)
+
{\int }_{Q_{-}}^{Q_{+}}
d{\Lambda' }\;2R_q(2(\Lambda-\Lambda'))
{\rho }({\Lambda' }),
\label{shiba1}
\end{equation}
where we have introduced the deformed Shiba-function\cite{shiba}
defined by
$
R_q(v)=\frac{1}{2\pi}
\sum_{m\in {\bf Z}}
\frac {e^{imv}}{1+q^{2|m|}}.
$
The energy is given by
\begin{equation}
E/L
=
2\ {\cosh {\gamma }}
-2{\pi }\ {\sinh {\gamma }}
\ {\biggl[ }\ 2R_q(0)
+
{\int }_{Q_{-}}^{Q_{+}}
d{\Lambda }\;2R_q(-2\Lambda )\;
{\rho }({\Lambda })\ {\biggr ]}.\label{eq:KY2}
\end{equation}
Thanks to
(\ref{shiba1}) and (\ref{eq:KY2}), we can
immediately apply the
general argument and  the results are:
i)
the charge sector can be described by the
$c=1$ bosonic conformal field theory
i.e. it
belongs to the universality class
called {\it the Tomonaga-Luttinger liquid},
ii)
the compactification radius $\cite {c1}$
is given by $r=\xi(Q)$,
where the {\it dressed charge} $\xi(\Lambda)$
satisfies
$
{\xi }(\Lambda)
 =1+
{\int }_{-Q}^{Q}
d{\Lambda' }\;
2R_q(2(\Lambda-\Lambda'))\ {\xi }({\eta }),
$
and  $Q$ is determined by
$
\left[{\int}_{-{\pi }/2}^{-Q}+{\int}_{Q}^{{\pi }/2}\right]
d\Lambda \;{{\rho }(\Lambda )}={\frac {N/2}{L}}.
$

As usual, we parameterize  $r$
by $K_{\rho }=r^{2}/2$. The equation for $\xi$  was solved numerically
and $K_{\rho}$ as functions of $n$ are shown in Fig. 6.
The relations between $K_{\rho }$ and the critical exponents
are shown in Table I \cite {hal,sg}.
As long as
$\gamma\neq 0$,
the singlet-superconducting correlation
is dominant
when $K_{\rho }>1$
in the low density region (high doping).
However for the usual supersymmetric case ($\gamma=0$), the superconducting
correlation can not be dominant in any filling \cite{ky} as seen in
Fig. 6.
This indicates that deformed $t$-$J$ models
including ours
may be more appropriate to study superconducting mechanisms
than the ordinary $t$-$J$ model.

The authors
benefited
from the help in the numerical calculations
by  Y. Hatsugai and Keiko Kimura.
J. S.
thanks
Kazuhiro Kimura for
discussions on the deformed $t$-$J$ model Hamiltonian.
They are also grateful to
M. Ogata and S.-K. Yang
for many useful comments and suggestions
on the algebraic Bethe ansatz and the finite-size scaling analysis.

\begin{figure}

Fig. 1.
Ground state energy per site
as a function of electron density $n$.
The solid lines are obtained by solving (\ref{intg}).
We also performed direct numerical diagonalizations of the Hamiltonian
for $L=14$
with $\gamma=2.0$, $1.0$ and $0.5$.
The results are plotted by
$\put (0.5,0.5){\framebox(4,5)}$
\ , + and ${\diamond }$ respectively.

Fig. 2.
Dispersion of the elementary excitations in the
charge sector for $\gamma=2$ and
$n=0.45$. Sufficiently many points
in the continuous spectrum are shown.
The momentum $P$ is periodic with period $2 \pi$.
The gapless points
are at $P=0,\ 2k_{F}$ and $2{\pi }-2k_{F}$.

Fig. 3.
Dispersion of the elementary excitations in the
spin sector for $\gamma=2$ and
$n=0.45$.
Sufficiently many points in the continuous spectrum are shown.
The momentum $P$ is periodic with period $2 \pi$.
There is no gapless point.
Note that the momentum $P$ for the elementary spin excitation
is  restrected in $[-P_{m},P_{m}]$, where
$P_{m}$ depends on $\gamma$ and $n$.

Fig. 4.
Spin gap as a function
of ${\gamma }$.
For $n=1$,
the spin gap of our model
reduces to
that of the XXZ spin chain
(see Eq.(\ref{ham1}))\cite {dp}.

Fig. 5.
Spin-spin correlations
on a logarithmic scale
for
the parity-unbroken Hamiltonian
${\cal H}_{tJ}-\sum_i \eta(S^z_i n_{i+1}-n_i S^z_{i+1})$.
The results are obtained by
numerical diagonalization of $L=12$ systems for
(a)\ $ N=10,\ \gamma =1.5$, (b)\ $ N=8,\ \gamma =1.5$
and (c)\ $ N=2,\ \gamma =2.0$.

Fig. 6
$K_{\rho }(n)$'s are shown for $\gamma=0.0,\;0.5,\;1.0$ and $2.0$.
The broken line (for $\gamma =0$ i.e. the ordinary supersymmetric case)
denotes the data from Ref. 8.
It can be shown analytically that
$K_{\rho}(0) =2$ and $K_{\rho}(1) =1/2$ for any $\gamma>0$.

\end{figure}

\vspace{10mm}

\begin{center}
\begin{minipage}{100mm}
\begin{table}
\caption{
Relation
between $K_{\rho }$ and the critical exponents of
the correlation functions.
}
\label{table}
\vspace{4mm}
\begin{tabular}{cc}
${\rm Correlations} $ & ${\rm exponents}$ \\
\tableline
${\rm 2k_{F}\ SDW \;\;(spin \;density \;wave)}$ & ${\rm exponential\ decay}$ \\
${\rm 2k_{F}\ CDW\;\;(charge \;density \;wave)}$ & $K_{\rho }$ \\
${\rm SS\;\;(singlet\;superconductivity)}$ & $1/K_{\rho }$ \\
${\rm TS\;\;(triplet\;superconductivity)}$ & ${\rm exponential\ decay}$ \\
${\rm 4k_{F}\  CDW\;\;(charge \;density \;wave)}$ & $4K_{\rho }$ \\
\end{tabular}
\end{table}
\end{minipage}
\end{center}

\begin{references}
\bibitem{XYham}
The Hamiltonian (\ref{ham1}) can be transformed to
that for the doped $XY$-model with broken parity:
$
{\cal H}_{XY}
=\sum_{i}\bigl[
 -\; \sum _{{\sigma }}
 {\cal P} (c_{i {\sigma }}^{\dagger }c_{i+1 {\sigma }}
     +c_{i+1 {\sigma }}^{\dagger }c_{i {\sigma }}){\cal P}
 +
2\;  (S_{i}^{x}S_{i+1}^{x}+S_{i}^{y}S_{i+1}^{y})
-(e^{\gamma}  \; n_{i\uparrow}n_{i+1\downarrow}
+e^{-\gamma}  \;n_{i\downarrow}n_{i+1\uparrow})
\bigr].
$
\bibitem{branch}
The branch of the logarithm is
fixed by the requirements that
$\phi(0,\alpha)=i \pi$, and ${\rm Im}\ \phi(x,\alpha)$ is a continuous
monotonic decreasing function in $-\pi/2<x<\pi/2$.
\end{references}
\end{document}